%% file: acl_latex.tex
\pdfoutput=1

\documentclass[11pt]{article}

\usepackage[final]{acl}
\usepackage{times}  
\usepackage{helvet}  
\usepackage{courier}  
\usepackage{natbib}  
\usepackage{caption} 
\usepackage{algorithm}
\usepackage{algorithmic}

\usepackage[utf8]{inputenc} 
\usepackage[T1]{fontenc}    
\usepackage{url}            
\usepackage{booktabs}       
\usepackage{amsfonts}       
\usepackage{nicefrac}       
\usepackage{microtype}      
\usepackage{xcolor}         
\usepackage{graphicx}
\usepackage{colortbl}   
\usepackage{multirow}  
\usepackage{diagbox}
\usepackage{amsmath}
\usepackage[most]{tcolorbox}
\tcbuselibrary{theorems}
\usepackage{subcaption} 
\usepackage{cuted}
\usepackage{textcomp}

\usepackage{tabularx}
\usepackage{enumitem}

\tcbset{
  mysectionbox/.style={
    colframe=blue,
    colback=white,
    boxrule=0.4pt,
    arc=0mm,
    left=1mm,
    right=1mm,
    top=0.5mm,
    bottom=0.5mm,
    fontupper=\scriptsize
  }
}

\usepackage{newfloat}
\usepackage{listings}
\DeclareCaptionStyle{ruled}{labelfont=normalfont,labelsep=colon,strut=off} 
\floatstyle{ruled}
\newfloat{listing}{tb}{lst}{}
\floatname{listing}{Listing}
\title{Sphinx: Benchmarking and Modeling for LLM-Driven Pull Request Review}

\author{
  \textbf{Daoan Zhang\textsuperscript{\textdagger,\textdaggerdbl}\thanks{Work done as an intern at Microsoft}}
  \quad
  \textbf{Shuo Zhang\textsuperscript{\textdaggerdbl}}
  \quad
  \textbf{Zijian Jin\textsuperscript{\textdaggerdbl}}
  \quad
  \textbf{Jiebo Luo\textsuperscript{\textdagger}} \\
  \quad
  \textbf{Shengyu Fu\textsuperscript{\textdaggerdbl}}
  \quad
  \textbf{Elsie Nallipogu\textsuperscript{\textdaggerdbl}}
  \\
  \textsuperscript{\textdagger}University of Rochester
  \quad
  \textsuperscript{\textdaggerdbl}Microsoft
  \\
  \small{\texttt{\{daoan.zhang,jluo\}@rochester.edu}} \\
  \small{\texttt{\{shuzha, zijianjin, shengyfu, elsie.nallipogu\}@microsoft.com}}
}

\begin{document}
\maketitle

\begin{abstract}
Pull request (PR) review is essential for ensuring software quality, yet automating this task remains challenging due to noisy supervision, limited contextual understanding, and inadequate evaluation metrics. We present \textbf{Sphinx}, a unified framework for LLM-based PR review that addresses these limitations through three key components: (1) a structured \textit{data generation pipeline} that produces context-rich, semantically grounded review comments by comparing pseudo-modified and merged code; (2) a \textit{checklist-based evaluation benchmark} that assesses review quality based on structured coverage of actionable verification points, moving beyond surface-level metrics like BLEU; and (3) \textit{Checklist Reward Policy Optimization (CRPO)}, a novel training paradigm that uses rule-based, interpretable rewards to align model behavior with real-world review practices. Extensive experiments show that models trained with Sphinx achieve state-of-the-art performance on review completeness and precision, outperforming both proprietary and open-source baselines by up to 40\% in checklist coverage. Together, Sphinx enables the development of PR review models that are not only fluent but also context-aware, technically precise, and practically deployable in real-world development workflows. The data will be released after review.
\end{abstract}

\section{Introduction}

Pull Request (PR) review~\cite{zhao2019improving, zhang2022pull, yang2025code} is a fundamental practice in collaborative software development, ensuring that proposed code changes are not only functionally correct but also maintainable, stylistically consistent, and aligned with broader project requirements. An effective PR review requires a deep understanding of the introduced diffs, the surrounding code context, the motivation and design intent captured in the PR description, and often the background discussed in linked issue threads. Reviewers synthesize these heterogeneous signals to produce precise, actionable feedback that improves code quality and reinforces good development practices.
While there has been growing interest in automating aspects of PR review—such as acceptance prediction and comment generation—the generation of \emph{high-quality, contextually grounded, and systematically evaluated} review comments remains underexplored. Existing datasets for this task typically rely on noisy, human-authored comments, which vary widely in relevance, coverage, and clarity. Moreover, many current methods treat PRs as isolated code diffs, overlooking the rich multi-modal context that is critical for meaningful review generation. Evaluation practices further exacerbate these issues by relying primarily on surface-level text similarity metrics such as BLEU~\cite{papineni2002bleu} and ROUGE~\cite{lin2004rouge}, which fail to reflect whether a review meaningfully identifies substantive issues in the proposed code changes.
In contrast, human reviewers often operate with implicit or explicit checklists—verifying functional correctness, robustness, security compliance, coding conventions, and architectural alignment. However, no existing benchmark or evaluation framework explicitly models this structured process. As a result, current automated systems tend to generate fluent but shallow comments, frequently missing critical issues or offering vague suggestions.

To address these challenges, we introduce \textbf{Sphinx}, a comprehensive framework for PR review comment generation. Rather than focusing solely on dataset scale, Sphinx emphasizes a structured \textit{data generation pipeline} that tightly grounds review comments in verifiable code modifications. We leverage LLMs to synthesize \textbf{pseudo pending solution} based on the PR's intent and compare it against the \textbf{actual merged code} to generate review comments aligned with semantic and structural differences. This enables consistent, interpretable training signals for learning PR reviewers.
Beyond data construction, we propose two complementary contributions. First, we introduce a \textbf{Checklist-based Evaluation Framework}, which transforms review outputs into structured checklists of verification points, allowing fine-grained evaluation of how well a model captures actual modification intent. Second, we propose \textbf{Checklist Reward Policy Optimization (CRPO)}, a novel training paradigm that leverages these structured rewards—augmented with LLM feedback and regularized by response length—to guide models toward producing comprehensive and precise review comments.
Together, Sphinx, Checklist-based Evaluation, and CRPO provide a principled foundation for advancing PR review automation—yielding models that are not only linguistically fluent, but also technically rigorous, context-aware, and practically applicable in real-world development workflows.

This work makes the following contributions:

\begin{itemize}
    \item We propose \textbf{Sphinx}, a comprehensive framework for PR review automation that combines a structured data generation pipeline, a task-specific evaluation benchmark, and a reward-driven training strategy. Rather than relying on human-authored comments, Sphinx uses LLMs to synthesize pseudo-modified code based on PR intent and compares it to the actual merged code to generate review comments that are tightly grounded in real code changes.

    \item We introduce a \textbf{Checklist-based Evaluation Framework}, which derives ground-truth verification items from structured diffs and evaluates whether model-generated comments address them. This protocol enables fine-grained, interpretable, and task-aligned assessment of review quality, going beyond surface-level text similarity.

    \item We propose \textbf{Checklist Reward Policy Optimization (CRPO)}, a novel training paradigm that leverages checklist-derived reward signals and length-aware regularization to guide models toward generating precise, concise, and contextually faithful review comments—without relying on free-form critiques.

    \item We conduct extensive experiments on the Sphinx Benchmark, showing that models trained with our framework substantially outperform both open-source and proprietary LLMs, achieving up to 40\% higher checklist coverage compared to strong baselines like GPT-4.1.
\end{itemize}

\section{Related Work}
Automated PR review has been widely studied due to its high cost, time demands, and the inherent subjectivity of human reviewers~\cite{yang2024survey, yu2024fine, lin2024improving, feng2020codebert, cihan2024automated, lu2025towards, sun2025bitsai}. A major challenge in this field lies in the quality of training data: most existing approaches rely on noisy, human-authored review comments that often lack clear grounding in the actual code changes, leading to ambiguous supervision signals~\cite{tufano2022using, frommgen2024resolving}. In addition, evaluation practices have largely depended on reference-based metrics such as BLEU and ROUGE-L\cite{li2022automating}, which emphasize surface-level textual similarity and fail to capture whether review comments are semantically accurate or actionable. While LLMs such as GPT series and LLaMa-Reviewer~\cite{lu2023llama} have recently been introduced and show improved performance through pretraining, their training objectives remain loosely aligned with real-world review tasks. These models~\cite{thongtanunam2022autotransform, sulun2021rstrace+, li2025codedoctor, naulty2025bugdar} often lack structured guidance and struggle to generate reviews that are precise and comprehensive, highlighting persistent limitations in data quality, evaluation fidelity, and training methodology.
To address these challenges, we propose Sphinx, a unified framework that advances automated code review by improving all three key components of the pipeline.

\section{The Sphinx Data Pipeline \& Benchmark}

Despite increasing interest in automating pull request (PR) review, existing datasets fall short in two critical ways. First, they predominantly rely on noisy, human-authored review comments, which often lack consistent grounding in the associated code changes, resulting in ambiguous supervision for learning models. Second, they treat PRs in isolation, ignoring the rich multi-modal context—such as PR descriptions, linked issue discussions, and surrounding code—that human reviewers routinely utilize. These limitations hinder the development of models that are capable of producing structured, contextually grounded, and practically useful review comments.

To train effective PR review models, inputs typically include PR metadata, the original code, and a pseudo solution (i.e., a code variant containing plausible, review-worthy mistakes), while the expected output is the corresponding review comments. However, only PR metadata and original code can be reliably scraped from GitHub at scale. In contrast, high-quality pseudo solutions and grounded PR reviews are not directly available from public repositories. To bridge this gap, we construct the \textbf{Sphinx Dataset \& Benchmark}, which is designed to tightly couple review comments with meaningful code modifications under realistic project contexts. This dataset provides the necessary structure and contextual richness to enable the development of models that generate high-quality, grounded, and actionable PR reviews.

\begin{figure*}
    \centering
    \includegraphics[width=\linewidth]{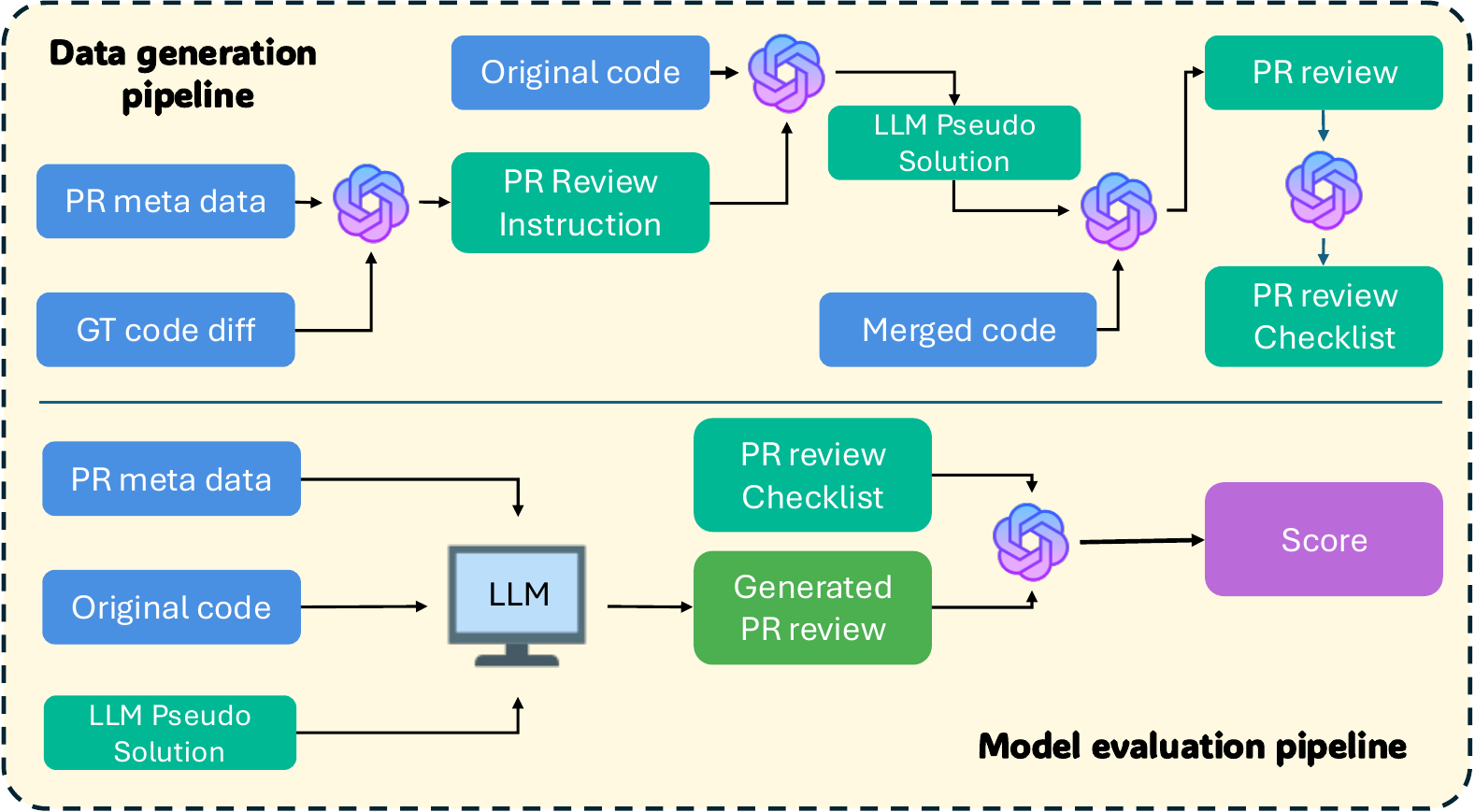}
    \caption{Overview of the PR review workflow, illustrating the transition from original code through generated solutions to final merged code, supported by a structured data generation and model evaluation pipeline.}
    \label{fig:enter-label}
\end{figure*}

\subsection{The Sphinx Data Pipeline}

In this section, we detail our methodology for collecting, processing, integrating, and generating data for PR review tasks. Our overarching objective is to construct a high-quality dataset tailored to the development and evaluation of LLM-based PR review models. To this end, we systematically designed a data pipeline that ensures each review instance is not only contextually grounded and semantically meaningful but also suitable for both supervised learning and structured evaluation. Our approach emphasizes data integrity, contextual richness, and alignment with real-world software development practices.

\paragraph{Data Collection.}
We systematically crawl pull requests from a curated set of popular, actively maintained GitHub repositories spanning five major programming languages: Java, JavaScript, C++, C\#, and Python. At this stage, we sampled approximately 200,000 PRs as the initial dataset.
For each pull request, we collect:
\begin{itemize}
    \item \textbf{GT code diff}: The groundtruth line-level code changes introduced by the PR.
    \item \textbf{PR meta data}: Natural language summaries authored by contributors describing the motivation and contents of the PR. Also including discussions, bug reports, or feature requests associated with the PR, providing additional background context.
    \item \textbf{Original code}: The original code before the PR was merged.
    \item \textbf{Merged code}: The final integrated version of the code after the PR is merged into the main branch.
\end{itemize}

\paragraph{Data Filtering.}
At this stage, we primarily perform data filtering and annotation to ensure the quality and consistency of the collected pull requests. Approximately 75,000 PRs were retained after filtering. We apply the following filtering criteria:

\begin{enumerate}
    \item \textbf{Completeness Check}: We verify that each crawled pull request contains all required components, including diffs, PR descriptions, linked issues, original code, and merged code.

    \item \textbf{Merged status}: Only pull requests that have been successfully merged into the main branch are included, as they represent reviewed and accepted contributions.
    
    \item \textbf{Input Length Constraint}: We retain only PRs that modify a single file in supported programming languages and discard those where the combined length of PR metadata and original code exceeds 32K tokens, ensuring fair comparison across most large language models. Support for multi-file changes and long-context data will be added in future versions.
    
    \item \textbf{Safety Filtering}: We apply GPT-4o-based safety screening mechanisms to remove PRs that potentially contain harmful, offensive, or otherwise unsafe content.
\end{enumerate}

\paragraph{LLM-Assisted PR review Generation.}

To construct high-quality PR reviews, we follow a structured process involving pseudo code generation and review synthesis:

\begin{enumerate}
    \item \textbf{PR Review Instruction Generation:} We prompt a LLM with the PR metadata and code diffs to generate a \textit{PR Review Instruction}, which captures the intent and objectives of the proposed code change.
    
    \item \textbf{LLM Pseudo Solution:} Given the original code and the PR Review Instruction, the LLM generates a pseudo version of the modified code that realizes the PR's intended functionality(Simulating a PR that requires review).
    
    \item \textbf{Comparison and Review Generation:} We compare the pseudo PR code with the actual merged code using the LLM, identifying semantic and structural differences. These discrepancies are then distilled into fine-grained, actionable review comments that are grounded in the intended change context.
\end{enumerate}

This structured approach ensures that the generated review comments are closely aligned with meaningful code changes, rather than superficial text edits. By grounding feedback in intentional discrepancies between LLM-generated and ground-truth implementations, our method produces reviews that are precise, actionable, and free from the noise typically found in human-authored PR comments. After this stage, approximately 45,000 PRs were retained.

\subsection{The Sphinx Benchmark Construction}

To construct the Sphinx benchmark, we randomly sample 500 data-points for each programming language represented in the Sphinx dataset (2500 in total for 5 languages). Among these, 450 buggy examples are selected, while the remaining 50 are confirmed to be bug-free. This sampling strategy allows the benchmark to evaluate both the ability to identify defects and the precision in recognizing correct code changes. Including bug-free examples reflects realistic PR review scenarios and helps prevent models from being overly biased toward negative (i.e., overly "mean") outputs.

\paragraph{PR Review Task Definition.}
Given the inputs—PR diffs, PR descriptions, linked issues, and code context—the benchmark defines the core task as generating high-quality PR review comments that (i) are grounded in the code changes, (ii) demonstrate contextual understanding of the PR's intent, and (iii) provide actionable, constructive feedback.

\paragraph{Ground-Truth Checklist Generation.}
From the structured diffs between modified and merged code, we automatically derive a \textbf{ground-truth checklist} for each PR. Each checklist item corresponds to a discrete verification point, such as enforcing consistent naming conventions, ensuring proper error handling, optimizing performance-critical paths, or adhering to project-specific APIs. This checklist serves as a structured representation of what a thorough reviewer would be expected to verify based on the actual code modifications.

\paragraph{Evaluation Protocol.}
Rather than relying solely on lexical overlap metrics, we adopt the Checklist-based Evaluation Framework. Specifically, for each PR, the ground-truth checklist provides the set of essential verification points. Then the model-generated review comments are assessed using an LLM to judge whether they adequately address each checklist item. We further define a weighted scoring function that considers both buggy (checklist-based) and bug-free (no-comment) cases, shown in Equation. ~\eqref{score}.


\begin{multline}
\label{score}
\text{Checklist-coverage Score} = 
\lambda \cdot \frac{1}{n} \sum_{i=1}^{n} \left(
 \frac{S^{(i)}_{\text{buggy}}}{N^{(i)}_{\text{buggy}}} \right) \\
+ (1 - \lambda) \cdot \frac{1}{m} \sum_{j=1}^{m} \left(
 \frac{S^{(j)}_{\text{bug-free}}}{N^{(j)}_{\text{bug-free}}} \right)
\end{multline}

Here, \(S^{(i)}_{\text{buggy}}\) and \(S^{(j)}_{\text{bug\text{-}free}}\) denote the checklist coverage counts for the \(i\)-th buggy case and the \(j\)-th bug-free case, respectively. \(N^{(i)}_{\text{buggy}}\) and \(N^{(j)}_{\text{bug\text{-}free}}\) represent the number of checklist items in each case. For bug-free cases, the checklist contains only one item: ``No comment.''

In our experiments, \(n\) is \(450\)  and \(m\) is \(50\) for each language and we set \(\lambda = 0.9\) to emphasize the contribution of checklist-based cases in the final metric.

\begin{figure*}[h!]
    \centering
    \includegraphics[width=\textwidth]{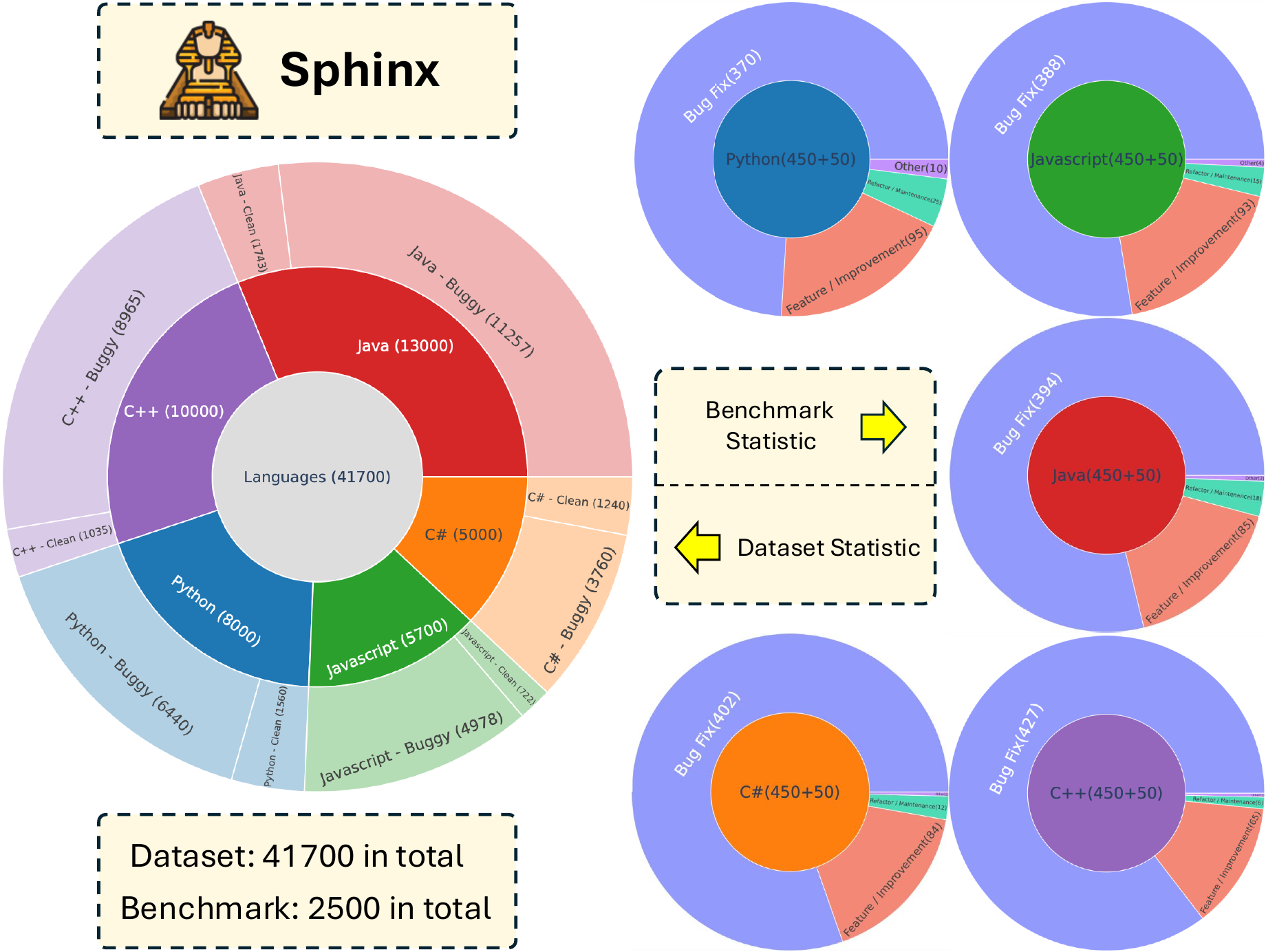}
    \caption{Data statistics of the Sphnix dataset and benchmark used for model training and evaluation. Details are in the Appendix.}
    \label{fig:stastic}
\end{figure*}

\subsection{Benchmark Statistics}

To support systematic evaluation, we construct a benchmark composed of 2,500 pull requests, spanning five widely-used programming languages: Python, JavaScript, Java, C\#, and C++. For each language, the benchmark includes 450 buggy cases and 50 bug-free cases, as visualized in Figure~\ref{fig:stastic}. The buggy cases contain realistic issues drawn from real-world development workflows, such as logic bugs, missing validations, API misuse, and naming inconsistencies. Each bug-free case corresponds to a clean code change, where the appropriate review response should be ``No comment.''
We categorize buggy samples into several semantic types, including:
\begin{itemize}
  \item \textbf{Bug Fix}: Addresses an error, defect, or unintended behavior in the codebase to restore correct functionality.
  
  \item \textbf{Feature / Improvement}: Introduces new functionality or enhances existing features, improving the capabilities or usability of the system.
  
  \item \textbf{Refactor / Maintenance}: Involves restructuring the code without altering its external behavior, such as code refactoring, dependency updates, or formatting adjustments.
  
  \item \textbf{Other}: Covers changes that do not clearly fall into the above categories, including documentation updates, build script modifications, or miscellaneous edits.
\end{itemize}

\subsection{Human Verification}

We conducted a survey to evaluate the reliability of our benchmark from three perspectives: Consistency between Review and Code Changes, Ground‑Truth Checklist Completeness \& Accuracy, and Practical Helpfulness \& Alignment with Developer Needs. We sampled 30 cases and collected responses from 20 reviewers. The resulting Mean Opinion Score (MOS) and Inter-rater Reliability (ICC(2,k)) were 4.13 and 0.87, respectively, demonstrating the effectiveness of our benchmark. Further details are provided in the Appendix.


\section{The Sphinx Model Training Pipeline}

We use 41,700 examples from the Sphinx dataset as our training set. We first apply supervised fine-tuning to adapt the base model to the PR review task. Then, we perform post-training to further enhance the model’s generalization ability on PR review.

\subsection{Supervised Finetuning (SFT) Stage}

In the supervised finetuning stage, we train the model using labeled data from the Sphinx dataset, where each training instance consists of PR code diffs, PR descriptions, linked issues, and the corresponding reference review comments. The goal of this stage is to align the model's output with high-quality, checklist-grounded review behaviors by minimizing the discrepancy between predicted and target comments.

\subsection{Post-training Stage}

Unlike code generation tasks with deterministic objectives, PR review is an open-ended task that lacks a rule-based reward function, making it especially prone to reward hacking when using heuristic evaluators such as LLM-as-a-judge~\cite{zheng2023judging}. In prior work, Constitutional AI~\cite{bai2022constitutional} addresses such alignment by generating rewards through critiques conditioned on a small set of \textbf{handcrafted principles}, formulated as:

\begin{equation}
\mathcal{R} = C \sim r_\theta \left( x, \{ y_j \}_{j=1}^{n}, \{ p_j \}_{j=1}^{m} \right),
\end{equation}

where \( x \) is the input (e.g., a code diff), \( \{ y_j \} \) are model responses, \( \{ p_j \} \) are guiding principles, \( C \) is a free-form critique, \( r_\theta \) is the reward function and \( \mathcal{R} \) is the resulting reward. While effective in general domains, this paradigm depends on generating textual critiques, which can be vague, inconsistent, or even exploitable.

To address these limitations in structured tasks like PR review, we propose replacing free-form critiques with \textit{structured checklists} derived directly from semantic code diffs. These checklists consist of concrete, verifiable items that serve as both evaluation anchors and discrete reward components. 
We formalize this approach as a new reward generation paradigm: \textbf{checklist-scalar}, extending the standard taxonomy of scalar, semi-scalar, and generative paradigms. Unlike prior methods that rely on free-form output, checklist-scalar defines reward as a function of checklist completion, offering consistent, interpretable, and domain-aligned supervision. Checklist-scalar removes the reliance on natural language outputs, reducing variance and mitigating reward hacking risks. This is particularly important in high-stakes settings such as code review, where vague or inconsistent critiques may lead to reward misalignment. Furthermore, because checklist items are predefined and task-specific, this paradigm enables reproducible reward computation and straightforward debugging—qualities that are critical for scaling LLM-based evaluation in structured domains.

While checklist-based reward brings strong interpretability and alignment benefits, it also introduces a potential vulnerability: longer review comments may have a higher likelihood of satisfying more checklist items, which can inadvertently encourage verbose outputs and lead to reward inflation. To mitigate this risk, we incorporate a smooth length-based penalty into the reward formulation. The penalty is defined based on the ratio between the predicted and reference checklist lengths—it remains zero when the prediction length is within \(M \times\) of the reference, and increases quadratically up to a cap when the length exceeds \(N \times\). This penalty is applied multiplicatively to the final reward, softly discouraging excessive verbosity while preserving flexibility for legitimate longer completions. This design helps maintain the precision and conciseness of generated reviews without destabilizing training.

The reward is defined as in \eqref{formula1}, where \( GT_i \) is the groundtruth code review, \( CL_i \) are the generated checklist items,  \( p_\theta \) is the checklist generation model and \( \gamma (|\{ y_i \}_{i=1}^{n}|, |\{ CL_i \}_{i=1}^{m}|) \) is a smooth penalty factor based on the ratio between the predicted length \( |\{ y_i \}_{i=1}^{n}| \) and the reference length \( |\{ CL_i \}_{i=1}^{m}| \). The penalty remains 1 when \( |\{ y_i \}_{i=1}^{n}| \leq M \cdot |\{ CL_i \}_{i=1}^{m}| \), and decays quadratically when \( |\{ y_i \}_{i=1}^{n}| > N \cdot |\{ CL_i \}_{i=1}^{m}| \), up to a capped minimum.


\begin{figure*}[t]
\centering
\begin{equation}
\label{formula1}
\begin{aligned}[t]
\{ CL_i \}_{i=1}^{m}
&\sim p_\theta \bigl( x, \{ GT_i \}_{i=1}^{n} \bigr), \\
\mathcal{R}
&= \gamma \bigl(
|\{ y_i \}_{i=1}^{n}|,\,
|\{ CL_i \}_{i=1}^{m}|
\bigr)
\cdot r_\theta \bigl(
x,\,
\{ y_i \}_{i=1}^{n},\,
\{ CL_i \}_{i=1}^{m}
\bigr)
\end{aligned}
\end{equation}
\end{figure*}


We refer to this framework as \textit{Checklist Reward Policy Optimization} (CRPO), an adaptation of  Group Relative Policy Optimization (GRPO)~\cite{shao2024deepseekmath} that integrates structured checklist rewards with calibrated length control. We also removed the KL loss in the original GRPO when training. CRPO provides consistent, interpretable, and robust reward signals without relying on free-form critiques, promoting precise and high-quality code review generation.

\section{Benchmarking and Experiment Results}

We conduct evaluations on a wide range of LLMs using our newly proposed Sphinx benchmark. The tested models include both commercial and open-source systems, including commercial models such as GPT-4.1, GPT-4o etc. On the open-source side, we evaluate models such as DeepSeek series, Qwen series etc. All the LLMs we used in the project is GPT-4o.
All evaluations follow the official inference settings released by the respective model developers, ensuring a fair comparison. We uniformly apply a standardized prompt template to maintain consistency across all evaluations.
In addition to evaluating off-the-shelf models, we further conduct supervised fine-tuning on both closed-source and open-source LLMs using the Sphinx dataset. Specifically, we fine-tune GPT-4o and GPT-4o-mini for the closed-source setting, and Qwen2.5-Coder-7B and Qwen2.5-Coder-14B for the open-source setting. Notice that, GPT-4o and GPT-4o-mini donot support GRPO training, so we just report the result after SFT. More details can be seen in the Appendix.

\subsection{Quantitative Results}

\definecolor{lightgray}{gray}{0.95}
\definecolor{lightblue}{RGB}{220,230,241}
\definecolor{lightgreen}{RGB}{220,240,220}
\definecolor{rowgray}{gray}{0.95}

\definecolor{headBlue}{HTML}{D0E3FA}
\definecolor{appleGray}{HTML}{FFF3E0} 
\definecolor{appleBlue}{HTML}{F3E5F5}
\definecolor{appleGreen}{HTML}{E9F7EF}

\begin{table*}[ht!]
    \centering
    \resizebox{\textwidth}{!}{
    \begin{tabular}{|c|c|c|c|c|c|c|c|c|}
    \toprule
    \rowcolor{headBlue}
    \multicolumn{1}{|c|}{} & \multicolumn{1}{c}{BLEU-1} & \multicolumn{1}{c}{ROUGE-L}  & \multicolumn{6}{c}{Checklist-coverage score} \\
    \hline
    \rowcolor{headBlue}
    \diagbox{Models}{Metrics} & Avg. & Avg. & JS & Java & C++ & Python & C\# & Avg. \\
    \hline
        \rowcolor{appleGreen}
        Starcoder~\cite{li2023starcoder}  & 1.64  & 2.60  & 14.53 &	13.56&	13.95&	16.28	&15.58	&14.78 \\
        \rowcolor{appleGreen}
        Starcoder2-7B~\cite{lozhkov2024starcoder}  & 1.78 & 2.56 & 15.35 &	18.13&	15.32	&18.97&	16.05	&16.76\\
        \rowcolor{appleGreen}
        Starcoder2-15B~\cite{lozhkov2024starcoder} & 1.65 & 2.65 &  17.34&	19.06	&22.67&	22.58&	19.28	&20.19\\
        \rowcolor{appleGreen}
        LLama3.3-70B~\cite{grattafiori2024llama} & 7.37 & 8.71 &  16.22	&17.32&	21.06&	22.81&	25.97&	20.68\\\rowcolor{appleGreen}
        Qwen2.5-Coder-7B~\cite{hui2024qwen2} & 10.51 & 12.04 &  14.29	&17.42	&26.64&	28.34&	25.83&	22.50\\\rowcolor{appleGreen}
        Qwen2.5-Coder-14B~\cite{hui2024qwen2} & 11.02  & 12.59 &  17.36&	20.05&	28.88&	30.72&	28.95&	25.19\\\rowcolor{appleGreen}
        Phi-4-14B~\cite{abdin2024phi} &  2.46& 3.33 &  21.40	&24.53	&26.90&	25.51&	27.80	&25.23\\\rowcolor{appleGreen}
        Deepseek-R1~\cite{guo2025deepseek} & 11.07 & 12.13 & 16.74& 20.72	&27.87&	32.60&	30.09&	25.60\\\rowcolor{appleGreen}
        Qwen2.5-Coder-32B~\cite{hui2024qwen2} & 11.74 & 12.06 & 22.32&	22.40&	29.03&	28.24	&26.17	&25.63\\\rowcolor{appleGreen}
        Qwen3-32B~\cite{qwen2.5} & 5.05 & 6.53 & 19.71	&19.47	&29.99&	32.15&	28.96&	26.06\\\rowcolor{appleGreen}
        DeepSeek-v3~\cite{liu2024deepseek} & 11.61 & 12.71 &16.94	&22.34&	28.58&	32.99	&31.39	&26.45\\\rowcolor{appleGreen}
        DSCoder-V2-L-Ins~\cite{zhu2024deepseek} & 11.40 & 11.70 & 22.31	&27.76&	22.79&	29.00&	38.57	&28.09\\\rowcolor{appleGreen}
        Qwen3-30B-A3B~\cite{qwen2.5} & 11.42 & 12.81 & 18.62 &	21.96&	34.02&	38.08&	33.03&	29.14\\\rowcolor{appleGreen}
        Qwen3-235B-A22B~\cite{qwen2.5} & 10.93 & 12.30 & 19.30&	23.36&	32.51&	38.88&	34.31&	29.67\\\rowcolor{appleGreen}
        Qwen2.5-72B-Ins~\cite{qwen2.5} & 11.52 & 12.22 & 21.42&	24.94&	33.10&	38.13&	35.06&	30.53\\\rowcolor{appleGray}
    \midrule
    GPT4.1-nano~\cite{openai2025gpt41}  & 10.90 & 12.32 & 15.35&	21.88&	31.48&	26.27&	32.14&	25.42\\\rowcolor{appleGray}
    GPT4.1-mini~\cite{openai2025gpt41}  & 11.56 & 12.64 & 15.59	&21.57&	31.38&	32.78	&32.38&	26.74\\\rowcolor{appleGray}
    GPT-4omini~\cite{openai_gpt4o_2024}  & 12.24 & 13.65 & 17.98	&22.56	&32.80&	36.73&	34.13&	28.84\\\rowcolor{appleGray}
    GPT4o-0806~\cite{openai_gpt4o_2024} & 12.35  & 13.38 & 18.23&	22.93	&34.39&	37.15&	34.32&	29.40\\\rowcolor{appleGray}
    Claude3.5-Sonnet~\cite{anthropic2024claude35} & 11.50 & 12.49 & 18.24&	23.07&	33.63&	36.95&	37.16&	29.81\\\rowcolor{appleGray}
    GPT-o3mini~\cite{openai2024gpt4omini} & 12.65  & 13.45 & 18.03	&23.84&	37.68&	39.73&	34.94&	30.84\\\rowcolor{appleGray}
    Claude3.7-Sonnet~\cite{anthropic2025claude37} & 11.40 & 12.25 &19.48	&25.92	&35.68	&38.20	&37.68	&31.39\\\rowcolor{appleGray}
    GPT-o4mini~\cite{openai2024gpt4omini} & 10.87 & 12.16 & 20.47&	23.82&	34.93&	40.81&	40.01&	32.01\\\rowcolor{appleGray}
    Gemini2.5-Pro~\cite{google2025gemini25pro} & 11.10  & 12.90 &21.30&	26.31&	38.59&	40.80&	36.75&	32.75\\\rowcolor{appleGray}
    GPT4.1~\cite{openai2025gpt41} & 11.50 & 12.52 & 21.82	&26.46&	41.85&	42.4&	38.62	&34.23\\\rowcolor{appleBlue}
    
    \midrule

    Sphinx-7B & 11.52 & 13.15 & 17.56&	23.80 &	28.34 &	29.43	&29.53 &25.73$^{\textcolor{red}{(+14.36\%)}}$
\\\rowcolor{appleBlue}
    Sphinx-14B & 11.86  & 13.65 &21.71 & 28.49&	33.32&	33.02&	34.54&	30.21$^{\textcolor{red}{(+19.92\%)}}$\\\rowcolor{appleBlue}
    Sphinx-4omini-SFT  & 12.74 & 14.76 & 29.74 &	36.84&	43.7&	41.87&	37.82	&37.99$^{\textcolor{red}{(+31.72\%)}}$\\\rowcolor{appleBlue}
    Sphinx-4o-SFT & 12.77 & 14.93 & 33.84 &	40.16	&44.84	&41.94&	44.83&	41.12$^{\textcolor{red}{(+39.86\%)}}$\\

    \bottomrule

    \end{tabular}}
    \caption{
\textbf{LLM performance on the Sphinx benchmark.} We report BLEU-1, ROUGE-L, and Checklist-coverage scores. Green rows show off-the-shelf models, orange rows are proprietary baselines, and purple rows are fine-tuned models. 
}
    \label{table:llm_performance}
\end{table*}

\subsubsection{Model Performance Analysis on the Sphinx Benchmark}

Table~\ref{table:llm_performance} summarizes the performance of a wide range of large language models (LLMs) on the Sphinx benchmark, evaluated across five programming languages (JavaScript, Java, C++, Python, and C\#). We report both traditional generation metrics (BLEU-1 and ROUGE-L) and our task-specific metric, the \textbf{Checklist Score}, which better reflects semantic correctness, review precision, and actionable coverage in the context of PR review.

\textbf{Off-the-shelf open models} (highlighted in green) exhibit moderate performance, with most achieving Checklist Scores in the range of 20--30. Among these, Qwen2-5.72B-Ins performs the best, attaining an average Checklist Score of 30.53, with particularly strong results in Python (38.13) and C\# (34.05). However, even the top open models struggle with complex review reasoning and consistency across all languages, indicating limitations in general-purpose instruction tuning for this specialized task.

\textbf{Proprietary LLMs} (highlighted in orange), including GPT-4, Claude 3, and Gemini, demonstrate significantly higher Checklist Scores across all languages. GPT-4.0 achieves the highest average among these at 34.23, followed by Claude 3.5 Sonnet (33.61) and Gemini 2.5 Pro (32.75). These results underscore the benefits of large-scale alignment and proprietary reinforcement strategies, though the closed nature of these models limits transparency and adaptability. In contrast, our fine-tuned Sphinx models achieve the best overall performance by a wide margin. {Sphinx-4o-SFT reaches an average Checklist Score of \textbf{41.12}, outperforming GPT-4o by \textbf{+6.89}. These improvements are consistent across all five languages, with notable gains in traditionally challenging domains like C++ (44.84) and C\# (45.83). The strong performance illustrates the effectiveness of our \textit{checklist-guided supervised finetuning and reward design}, which provides the model with fine-grained, interpretable supervision signals tailored to real-world review practices.

Interestingly, although BLEU-1 and ROUGE-L scores remain relatively close across many models, the Checklist Score exhibits much higher variance and discriminative power. For instance, while GPT-4 and Sphinx-40-SFT have similar ROUGE-L scores (around 14.9), their Checklist Scores differ substantially (34.23 vs. 41.12). This reinforces our earlier claim that \textbf{BLEU and ROUGE do not adequately capture the semantic completeness or technical accuracy required in PR review}.


\begin{table}[htbp]
\centering
\small
\begin{tabular}{|c|c|c|c|}
\toprule

\rowcolor{headBlue}
\diagbox{Models}{Metrics} & BLEU-1 & ROUGE-L & CLC\\
\hline
\toprule

\rowcolor{appleGreen}
Qwen2.5-Coder-7B & 10.51 & 12.04 & 22.50 \\
\rowcolor{appleBlue}
Sphinx-7B-SFT & 11.89 & 13.08 & 24.10 \\
\rowcolor{appleBlue}
Sphinx-7B-SFT-CRPO & 11.52 & 13.15 & 25.73 \\
\midrule
\rowcolor{appleGreen}
Qwen2.5-Coder-14B & 11.02  & 12.59 & 25.19 \\
\rowcolor{appleBlue}
Sphinx-14B-SFT & 11.67  & 13.15 & 28.38 \\
\rowcolor{appleBlue}
Sphinx-14B-SFT-CRPO & 11.86  & 13.65 & 30.21 \\
\rowcolor{appleBlue}
\bottomrule
\end{tabular}
\caption{
\textbf{SFT \& CRPO Performances.} We report BLEU-1, ROUGE-L, and Checklist-coverage scores. Both SFT and CRPO can bring performance gains.
}
\label{tab:compare1}
\end{table}

\subsubsection{Effectiveness of Checklist-Based Training and Optimization}

As shown in Table~\ref{tab:compare1}, both supervised finetuning (SFT) and CRPO contribute to consistent performance improvements. The checklist score increases by up to +5.02 points for the 14B model (from 25.19 to 30.21), highlighting the effectiveness of checklist-aligned reward signals.
To better understand the learning dynamics, we visualize the mean critic reward and response length during CRPO training in Figure~\ref{fig:combined}. The left plot shows a steady increase in reward, indicating that the model is learning to generate outputs better aligned with checklist objectives. Meanwhile, the average response length remains relatively stable, suggesting that the reward improvement is not simply due to longer responses but rather more targeted and semantically relevant content. These results demonstrate that our rule-based reward not only guides the model toward better checklist coverage but also promotes stable and efficient learning.

\begin{figure}[htbp]
  \centering
  \begin{subfigure}[b]{0.4\textwidth}
    \includegraphics[width=\textwidth]{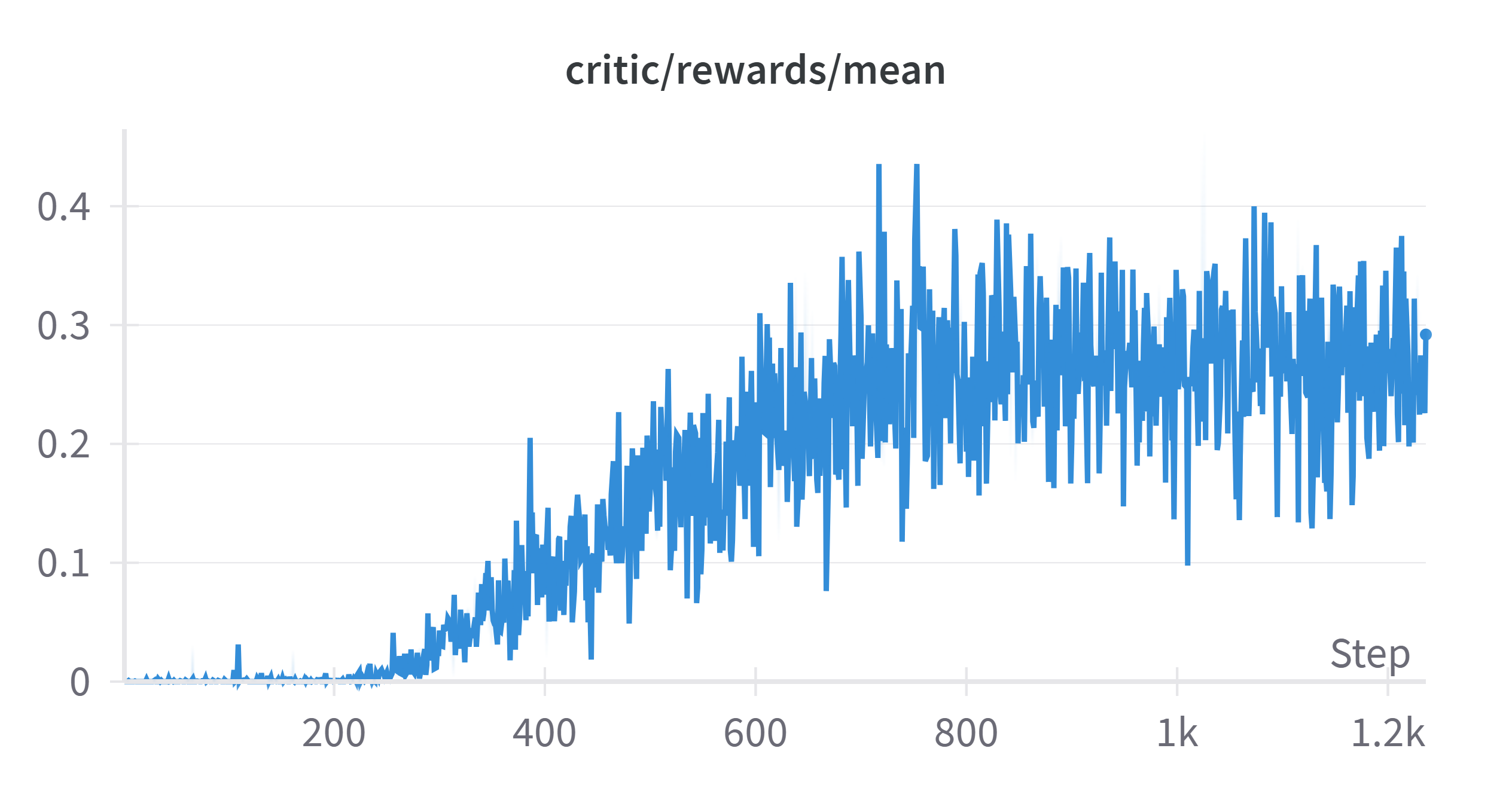}
  \end{subfigure}
  \hfill
  \begin{subfigure}[b]{0.4\textwidth}
    \includegraphics[width=\textwidth]{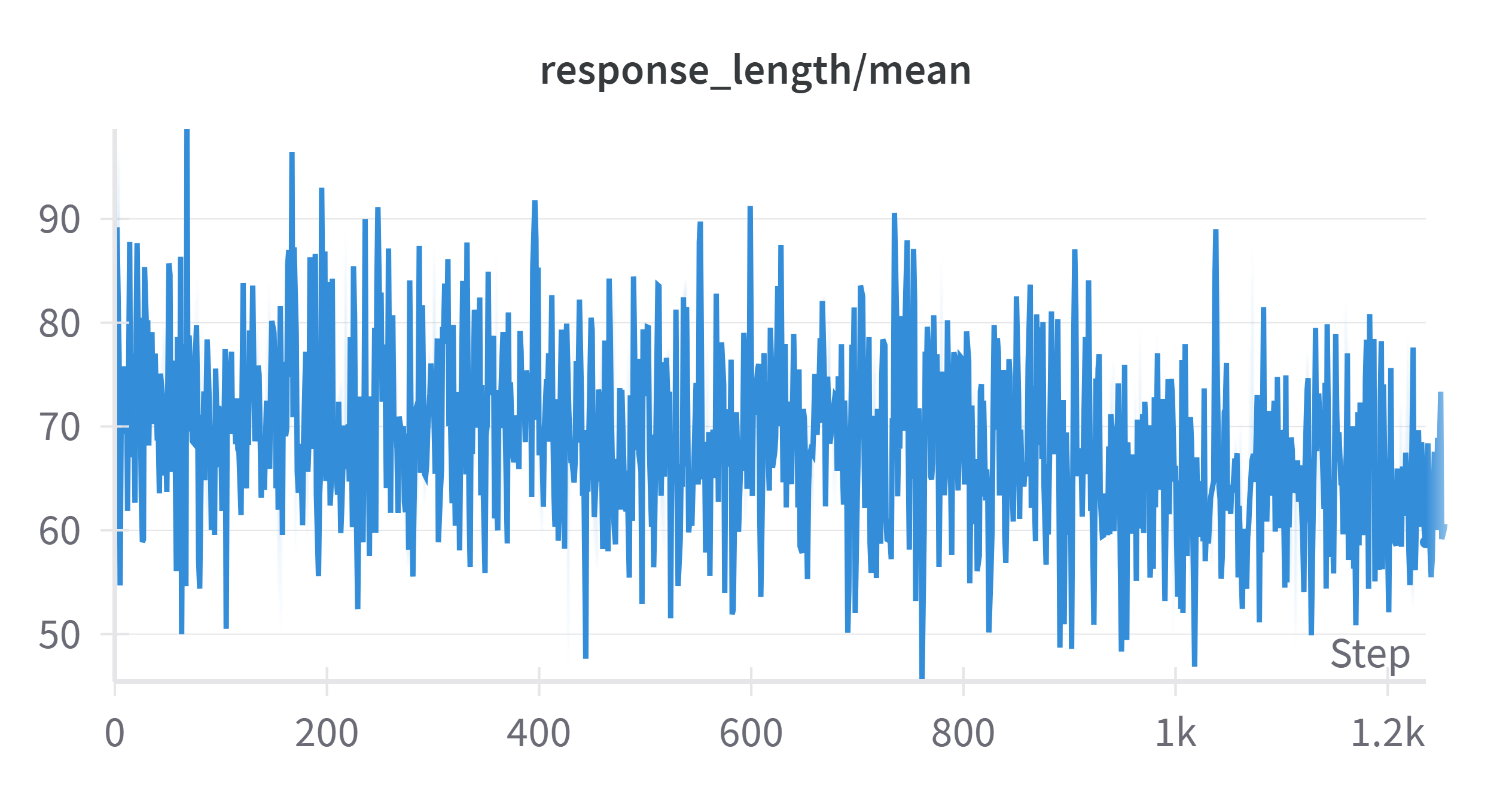}
  \end{subfigure}
  \caption{Reward(Left)/Response Length(Right) variation over steps.}
  \label{fig:combined}
\end{figure}

\begin{table}[htbp]
\centering
\small
\begin{tabular}{|c|c|c|c|}
\toprule

\rowcolor{headBlue}
\diagbox{Models}{Metrics} & BLEU-1 & ROUGE-L & CLC Score \\
\hline
\toprule

\rowcolor{appleGreen}
LLM-as-a-judge & 11.07 & 12.24 & 25.12 \\
\rowcolor{appleGreen}
No length penalty & 3.25 & 4.86 & 25.35 \\
\rowcolor{appleGreen}
With KL loss & 10.78 & 12.35 & 22.86 \\
\midrule
\rowcolor{appleBlue}
Sphinx-7B & 11.52 & 13.15 & 25.73 \\

\bottomrule
\end{tabular}
\caption{
\textbf{Ablations when using the Sphinx dataset.} We report BLEU-1, ROUGE-L, and Checklist-coverage scores. Both SFT and CRPO can bring performance gains.
}
\label{tab:compare}
\end{table}

\subsection{Ablation study}
To better understand the contribution of various training components in our pipeline, we conduct three ablation studies. As shown in Table.~\ref{tab:compare}, we firstly replace the checklist-based reward with an LLM-as-a-judge~\cite{zheng2023judging} reward. While this approach yields reasonable performance, it still underperforms our full model, suggesting that general-purpose reward models may lack the necessary task-specific signal for structured code review generation. Second, we evaluate the impact of removing the length penalty from the CRPO reward. Without this regularization, BLEU and ROUGE scores degrade severely, despite a comparable checklist coverage. This indicates that the model learns to game the reward by generating excessively verbose outputs that superficially match checklist terms. Lastly, we reintroduce a KL divergence loss during CRPO training which  leads to consistent drops across all metrics, suggesting that the base model may be not well aligned with the data distribution in our Sphinx dataset. Overall, these results demonstrate that each component—checklist-guided reward, length regularization, and training objective—plays an essential role in achieving high-quality, concise, and targeted code review generation.

\subsection{Qualitative Analysis}

The example~\ref{compare} shows that Sphinx-14B achieves complete checklist coverage, correctly handling both conditional credential setting and clearing logic. In contrast, Qwen2.5-14B-instruct omits key checks, such as verifying field presence and using \texttt{elif} to avoid overwriting existing values. Coverage is visually indicated with \textcolor{green!50!black}{green} for matched items and \textcolor{red}{red} for missing ones.

\begin{tcolorbox}[colframe=gray, colback=white, title=Checklist Coverage: Qwen2.5-Coder2.5-14B-instruct vs. Sphinx-14B, coltitle=white, colbacktitle=gray, fonttitle=\bfseries, fontupper=\scriptsize]

  \begin{tcolorbox}[mysectionbox, title=Groundtruth checklist]
\texttt{[\\
\ \ \ "Refactor authentication check to verify both mail\_useauth and at least one credential (mail\_u or mail\_p) exist (ensures credentials are only set when provided)",\\
\ \ \ "Update credential clearing logic to use an elif clause when mail\_useauth is None (preserves existing credentials when an empty value is intentionally submitted)"\\
]}
  \end{tcolorbox}

  \begin{tcolorbox}[mysectionbox, title=Qwen2.5-Coder2.5-14B-instruct (Checklist Coverage Score:0.5)]
...\\
\colorbox{green!20}{\textcolor{black}{Sets credentials if mail\_useauth is checked.}}
\colorbox{green!20}{\textcolor{black}{Clears them if unchecked.}} \\
\textbf{Missing:}
\colorbox{red!20}{\textcolor{black}{No check for whether mail\_u or mail\_p is present.}}
\colorbox{red!20}{\textcolor{black}{No elif logic for preserving existing credentials.}}\\
...
  \end{tcolorbox}

  \begin{tcolorbox}[mysectionbox,  title=Sphinx-14B (Checklist Coverage Score:1.0)]
...\\
\colorbox{green!20}{\textcolor{black}{• Handles mail\_useauth=None by clearing credentials.}}
\colorbox{green!20}{\textcolor{black}{• Checks if mail\_u or mail\_p exists before setting.}}
\colorbox{green!20}{\textcolor{black}{• Uses elif to avoid overwriting existing values.}}\\
...
  \end{tcolorbox}

\label{compare}
\end{tcolorbox}


\section{Conclusion}
We introduce \textbf{Sphinx}, a comprehensive framework for LLM-based PR review, including a large-scale dataset and benchmark, task-aligned evaluation metric, and reward-driven training method. By combining supervised finetuning with checklist-guided reward optimization, our models significantly outperform both open and proprietary baselines on the Sphinx benchmark. These results highlight the value of structured supervision and interpretable reward design for open-ended code review tasks. Future work will explore multi-file PRs, human-in-the-loop evaluation, and broader applications in collaborative software development.

\bibliography{custom}

\newpage
\appendix

\input{appendix}

\end{document}

%% file: appendix.tex
\tcbset{
  mysectionbox/.style={
    colframe=blue,
    colback=white,
    boxrule=0.4pt,
    arc=0mm,
    left=1mm,
    right=1mm,
    top=0.5mm,
    bottom=0.5mm,
    fontupper=\scriptsize
  }
}

%
\lstset{%
	basicstyle={\footnotesize\ttfamily},
	numbers=left,numberstyle=\footnotesize,xleftmargin=2em,
	aboveskip=0pt,belowskip=0pt,%
	showstringspaces=false,tabsize=2,breaklines=true}
\floatstyle{ruled}
\newfloat{listing}{tb}{lst}{}
\floatname{listing}{Listing}
%
\pdfinfo{
/TemplateVersion (2026.1)
}

\setcounter{secnumdepth}{0} 

\title{Sphinx: Benchmarking and Modeling for LLM-Driven Pull Request Review}



\section{Broader Impacts}

Sphinx can positively influence software quality by making code review automation more accessible, especially in large-scale or open-source settings. It may also shift human roles in review workflows toward higher-level reasoning and mentorship. However, there are risks: over-reliance on automated suggestions could suppress diverse programming practices or critical thinking. Additionally, if left unchecked, the models may propagate biases present in training data, such as reinforcing specific coding conventions or overlooking non-standard but valid solutions. Ensuring transparency, human oversight, and responsible deployment will be essential as such tools gain adoption.

\section{Data Statistics}

This section provides a more granular breakdown of our 2,500-sample evaluation benchmark, supplementing the statistics presented in the main body of the paper. The benchmark samples were meticulously curated from a larger corpus of real-world pull requests to ensure a balanced, diverse, and challenging evaluation set across five major programming languages.

As stated in the main text, each of the five languages contributes 500 samples to the benchmark. These are divided into 450 ``buggy'' cases, which require a code modification, and 50 ``bug-free'' cases, which serve as a negative control where no change is needed. The 450 buggy samples for each language are further broken down into four semantic categories to represent a wide spectrum of common development tasks.

The specific distribution for each language is as follows:

\begin{itemize}
    \item \textbf{Java}: Contains 500 total samples, composed of 50 bug-free cases and 450 buggy cases. The buggy cases are categorized into:
    \begin{itemize}
        \item \texttt{Bug Fix}: 394 samples
        \item \texttt{Feature / Improvement}: 25 samples
        \item \texttt{Refactor / Maintenance}: 21 samples
        \item \texttt{Other}: 10 samples
    \end{itemize}

    \item \textbf{C++}: Contains 500 total samples, composed of 50 bug-free cases and 450 buggy cases. The buggy cases are categorized into:
    \begin{itemize}
        \item \texttt{Bug Fix}: 427 samples
        \item \texttt{Feature / Improvement}: 13 samples
        \item \texttt{Refactor / Maintenance}: 0 samples
        \item \texttt{Other}: 10 samples
    \end{itemize}

    \item \textbf{C\#}: Contains 500 total samples, composed of 50 bug-free cases and 450 buggy cases. The buggy cases are categorized into:
    \begin{itemize}
        \item \texttt{Bug Fix}: 402 samples
        \item \texttt{Feature / Improvement}: 24 samples
        \item \texttt{Refactor / Maintenance}: 10 samples
        \item \texttt{Other}: 14 samples
    \end{itemize}

    \item \textbf{Python}: Contains 500 total samples, composed of 50 bug-free cases and 450 buggy cases. The buggy cases are categorized into:
    \begin{itemize}
        \item \texttt{Bug Fix}: 370 samples
        \item \texttt{Feature / Improvement}: 25 samples
        \item \texttt{Refactor / Maintenance}: 39 samples
        \item \texttt{Other}: 16 samples
    \end{itemize}

    \item \textbf{JavaScript}: Contains 500 total samples, composed of 50 bug-free cases and 450 buggy cases. The buggy cases are categorized into:
    \begin{itemize}
        \item \texttt{Bug Fix}: 388 samples
        \item \texttt{Feature / Improvement}: 28 samples
        \item \texttt{Refactor / Maintenance}: 22 samples
        \item \texttt{Other}: 12 samples
    \end{itemize}
\end{itemize}

This detailed breakdown shows the benchmark is heavily weighted towards the \texttt{Bug Fix} category (totaling 1,981 samples), which aligns with the primary goal of evaluating models on automated program repair. The inclusion of \texttt{Feature / Improvement} (115 samples), \texttt{Refactor / Maintenance} (92 samples), and \texttt{Other} (62 samples) tasks ensures that models are also tested on their ability to comprehend code intent beyond simple error correction. Finally, the 250 \texttt{bug-free} samples are a crucial component for measuring a model's precision and its ability to avoid generating false positives on correct code.

\begin{figure*}[h!]
    \centering
    \includegraphics[width=\textwidth]{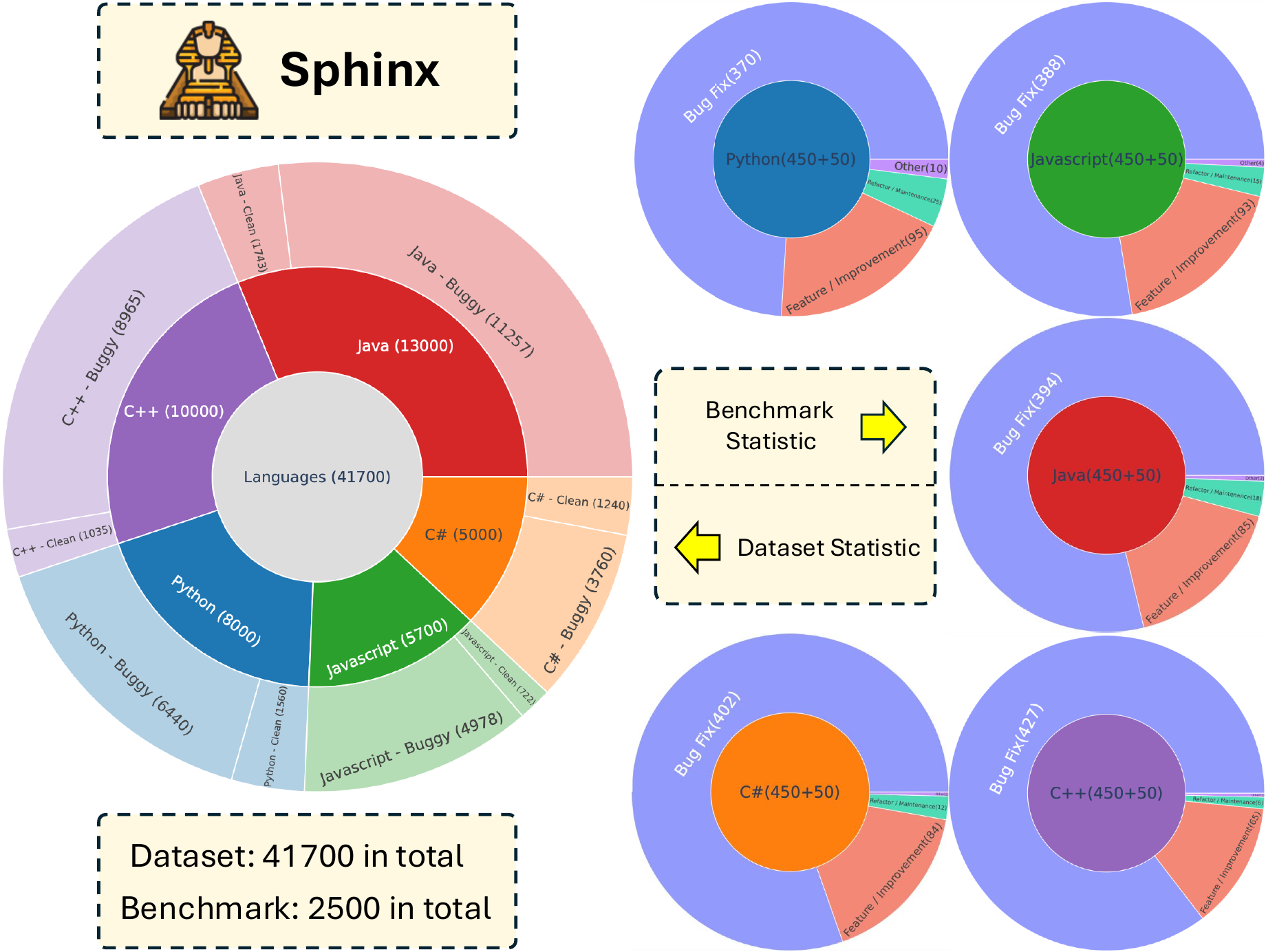}
    \caption{Data statistics of the Sphnix dataset and benchmark used for model training and evaluation. Details are in the Appendix.}
    \label{fig:stastic}
\end{figure*}


\section{Further Analysis}

Across the spectrum of tested models, from open-source systems to top-tier proprietary APIs, a distinct performance hierarchy among languages emerges. Models generally demonstrate higher proficiency when reviewing code written in Python and C\#. This trend suggests that the syntactic structure of these languages, or perhaps the sheer volume and quality of their representation in pre-training corpora, makes their review tasks more tractable for contemporary model architectures.

In contrast, languages such as C++ and Java consistently prove to be more challenging for the majority of the models. The performance gap between a model's score in Python versus its score in C++ is often substantial. This indicates that language features like manual memory management, complex type systems, and intricate template metaprogramming in C++ present a higher barrier for automated review, demanding a more profound level of code comprehension.

Notably, our Sphinx fine-tuning process appears to effectively mitigate this performance disparity. While our models exhibit improved performance across all languages, the gains are particularly pronounced in the traditionally more difficult languages like C++ and Java. This observation supports the conclusion that our targeted training methodology not only enhances general code review capabilities but is also particularly successful at instilling the nuanced understanding required to navigate the complexities of these more challenging programming domains.

\section{Training details}
\label{666}

\subsection{Supervised finetuning}

We use all the 41k data for training. All the 7B and 14B models were trained via lora with rank set to 8. The initial learning rate is 5e-5 with cosine learning rate schedule. The warm up step is set to 50. The batch size is set to 1024. We train for 165 steps for both of the two models.

\subsection{CRPO training}

We use GRPO as our base training procedure and change the reward to the checklist reward. We adopted the code in code-r1~\cite{code-r1}. The reward model used in all our models is o3mini. The learning rate is set to 1e-6 and the roll-out is set to 16. The gradient accumulation step is set to 4. Other hypermeters are the same with those in code-r1.

\section{Human Verification}

We used a questionnaire to assess the reliability of the content. We sampled 20 cases and collected responses from 20 reviewers from software engineering field. The questionnaire we used is as follows:

\subsection*{A. Consistency between Review and Code Changes}

\begin{tabularx}{\linewidth}{@{} l X c @{}}
\toprule
\textbf{ID} & \textbf{Statement} & \textbf{Rating (1--5)} \\
\midrule
A1 & The review accurately describes the functional changes introduced by the PR. & \\
\addlinespace
A2 & The review addresses all major code modifications shown in the diff. & \\
\addlinespace
A3 & Each review comment is mapped to the correct line(s) or section(s) of code. & \\
\addlinespace
A4 & Suggestions or critiques in the review are actionable given the current code. & \\
\bottomrule
\end{tabularx}

\vspace{6mm} 

\subsection*{B. Ground-Truth Checklist Completeness \& Accuracy}

\begin{tabularx}{\linewidth}{@{} l X c @{}}
\toprule
\textbf{ID} & \textbf{Statement} & \textbf{Rating (1--5)} \\
\midrule
B1 & The checklist covers all critical aspects that should be reviewed for this PR. & \\
\addlinespace
B2 & No irrelevant items are present in the checklist. & \\
\addlinespace
B3 & Wording of checklist items is clear and unambiguous. & \\
\addlinespace
B4 & Checklist items align with the issues actually present in the code diff. & \\
\bottomrule
\end{tabularx}



\subsection*{C. Practical Helpfulness \& Alignment with Developer Needs}

\begin{tabularx}{\linewidth}{@{} l X c @{}}
\toprule
\textbf{ID} & \textbf{Statement} & \textbf{Rating (1--5)} \\
\midrule
C1 & The generated review would help me improve or merge this PR more efficiently. & \\
\addlinespace
C2 & The checklist reflects my real-world concerns when performing code reviews. & \\
\addlinespace
C3 & I would trust this review \& checklist system in my daily workflow. & \\
\addlinespace
C4 & Overall, the combination of review and checklist adds value over a manual review alone. & \\
\bottomrule
\end{tabularx}

\vspace{6mm}
We further presented the distributions of all ratings and a heatmap of the ratings for the first 10 PRs.

\begin{figure*}
    \centering
    \includegraphics[width=0.8\linewidth]{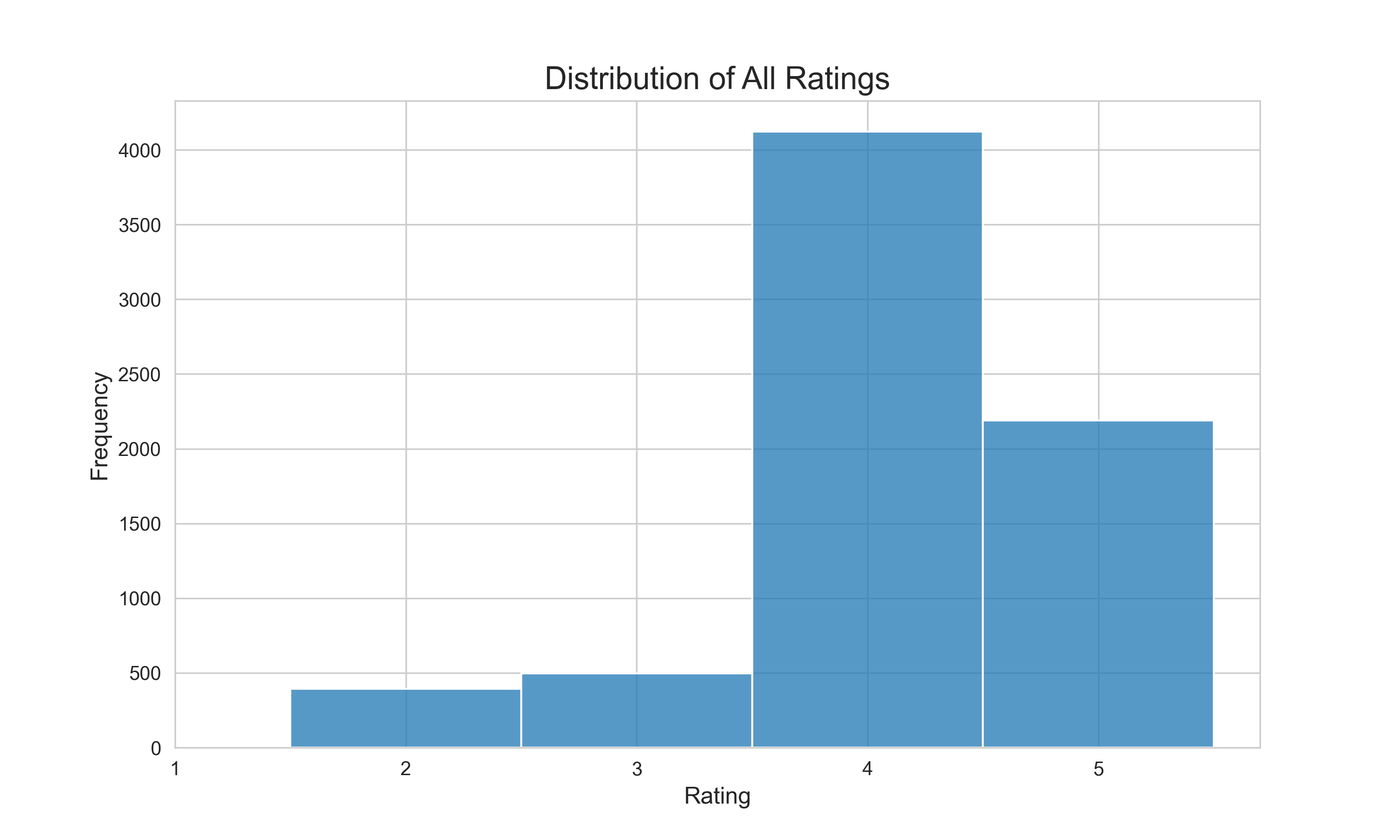}
    \caption{Rating Distributions}
    \label{fig:000}
\end{figure*}

As shown in Figure.~\ref{fig:000}, the rating distribution is strongly skewed towards the positive end, indicating high overall satisfaction. The most frequent rating is 4, which, combined with the second most frequent rating of 5, accounts for the vast majority of the data. Lower scores (1-3) are rare. This pattern suggests that while the quality was consistently perceived as high, raters were conservative, preferring to give a "very good" (4) over a "perfect" (5) score.

\begin{figure*}
    \centering
    \includegraphics[width=0.8\linewidth]{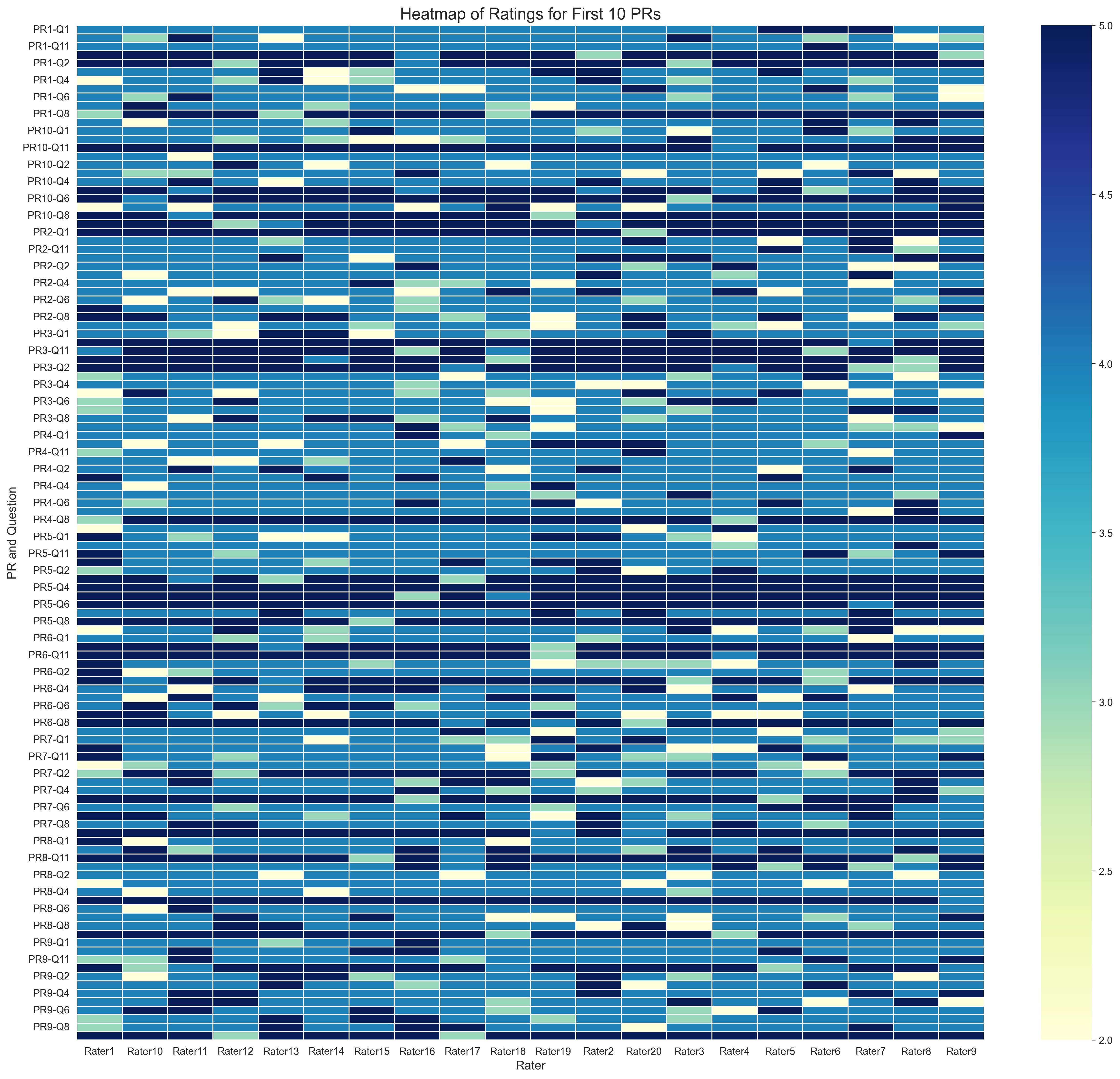}
    \caption{Rating Heatmaps for the first 10 PRs}
    \label{fig:001}
\end{figure*}

The rating heapmap is shown in Figure~\ref{fig:001}, this heatmap illustrates the ratings given by different raters (Rater1 to Rater20) for each question (Q1–Q12) across the first 10 PRs (PR1 to PR10). The abundance of dark blue areas indicates that ratings are mostly concentrated near the high end (close to 5.0), suggesting strong agreement among raters. Therefore, our benchmark shows a high degree of alignment with human judgment.

\section{Prompts used in Sphinx}

In this section, we present the full set of prompts used throughout the Sphinx project. These prompts are designed to guide LLMs in various code-related tasks, including instruction generation, code synthesis, PR review, checklist creation, and evaluation. Each prompt is carefully crafted to ensure consistency, clarity, and high-quality output across different components of the system. The prompts are provided in full below for reference and reproducibility.

\begin{tcolorbox}[colframe=gray, colback=white, title=PR review instruction generation prompt, coltitle=white, colbacktitle=gray, fonttitle=\bfseries]

    Problem Definition:
    
    Clearly summarize the issue or need for change based on the provided problem statement, related issues, and diff code.
    Code Editing Requirement: Provide a precise and detailed specification of the necessary code modifications.

    Guidelines:
    
    - Focus on code modifications only: Ignore changes related to other file types (e.g., HTML, CSS, Markdown, configuration files).
    
    - Extract Key Code Elements: Identify function names, variable names, classes, methods, etc., from Diff Format Code Changes to ensure precise and well-defined modifications.
    
    - Be specific and precise: Based on the Diff Format Code Changes, clearly define the modification target (function names/methods/variables), modification logic, and input-output requirements if applicable.
    
    - No business context or team discussions: Describe only the necessary code changes.

    Input:
    
    Problem Statement:
    
    \{ps metadata\}
    
    Related Issues: 
    
    \{issue\_info\}
    
    Diff Format Code Changes: 
    
    \{GT code\_diff\}

    Output json format:
    
    [{{"problem\_definition": 
        
        "[Concise explanation of the issue or requirement]",
   
        "code\_editing\_requirement": 
        
        [
        {{
            "modification\_target": "string",
            "modification\_logic": "string",
        }}
        ]
    }}]

\label{compare1}
\end{tcolorbox}

\begin{tcolorbox}[colframe=gray, colback=white, title=Prompt for LLM-generated code, coltitle=white, colbacktitle=gray, fonttitle=\bfseries]

You will be provided with a partial code base and an issue statement explaining a problem to resolve.

<issue>

PR review instruction

</issue>

<code>

Original code

</code>

 I need you to solve this issue by regenerating the full file content thatyou would like to change. Please respond with the complete edited code. Do not include explanations, comments, or any additional text.

\label{compare2}
\end{tcolorbox}

\begin{tcolorbox}[colframe=gray, colback=white, title=Prompt for PR review generation, coltitle=white, colbacktitle=gray, fonttitle=\bfseries, ]

You are presented with two versions of code:

Generated Version: a pseudo or automatically generated implementation (e.g., from an AI model).

Reference Version: a known correct or ground-truth implementation.

Your job is to analyze the differences between these two versions and provide pull request-style comments on the generated version.

**Generated Version**: 

\{LLM-generated code\}

**Reference Version**:

\{Merged code\}

**Your responsibilities:**

Point out mistakes, logical errors, or inefficiencies in the generated version.

Reference how the reference version handles the issue more correctly or efficiently.

Offer specific, actionable suggestions for how the generated version can be improved.

**Review Guidelines:**

Do not explain the full logic of the reference version—focus only on how it improves upon the generated version.

Frame feedback as PR comments directed at the generated code.

Be specific and objective. Avoid vague suggestions.

If both versions are correct but use different approaches, discuss trade-offs (e.g., performance, readability, maintainability).

If both versions are functionally equivalent and maintain similar code quality, output "No comment."

If the generated version omits important logic or uses incorrect assumptions, highlight these clearly.

Output your feedback as a list of PR comments, each attached to a specific problematic pattern or line in the generated code.

\label{compare3}
\end{tcolorbox}

\begin{tcolorbox}[
  breakable,
  colframe=gray,
  colback=white,
  title=PR review checklist generation,
  coltitle=white,
  colbacktitle=gray,
  fonttitle=\bfseries
]

Based on the following code reviews, generate a comprehensive and highly detailed checklist in plain list format. The checklist should cover all the issues, suggestions, and areas for improvement mentioned in the code reviews.

Checklist Requirements:

- Output as a plain list of strings (e.g., ["Ensure X", "Avoid Y", ...])
- Each checklist item must be a single string

- Start each item with an **action verb** (e.g., "Ensure", "Avoid", "Refactor", "Verify")

- Break down each review comment into the **smallest actionable step possible**

- Include a short explanation inside each string in parentheses to clarify the intent

- Be specific and avoid vague instructions (e.g., "Improve performance" → "Replace nested loops with hash map lookups for faster access (constant time instead of linear time)")

- Ensure the checklist fully reflects all the points raised in the review, covering at minimum:

   \hspace*{2em}   - Readability and clarity
    
   \hspace*{2em}   - Performance and efficiency
    
   \hspace*{2em}   - Code maintainability
    
   \hspace*{2em}   - Security concerns
    
   \hspace*{2em}  - Error handling and logging
    
   \hspace*{2em}  - Naming conventions and consistency
    
   \hspace*{2em}  - Framework/library usage
    
   \hspace*{2em}  - Tests and validation

Use the following sample code review to understand how to generate the checklist:

"""

1. Variable names like `a` and `b` are not meaningful.

2. There is a hardcoded value `60` used multiple times.

3. The `if` condition in line 20 is too deeply nested.

4. No error handling around file reading.

5. There is no test for when the input file is empty.

"""

Sample Output:

[

   \hspace*{2em} "Rename variables like 'a' and 'b' to descriptive names (improves readability and maintainability)",
    
  \hspace*{2em}  "Replace hardcoded value '60' with a named constant (avoids magic numbers and improves reusability)",
    
   \hspace*{2em} "Refactor deeply nested if-condition into a separate function (enhances clarity and modularity)",
    
   \hspace*{2em} "Add error handling around file reading (prevents crash on file access failures)",
    
    \hspace*{2em} "Add a test case for empty input files (ensures edge case is properly handled)"
    
]

Now use the following actual code review to generate the checklist:

{PR review}

Return only the checklist as a plain list. Do not add any explanation or formatting.

\label{compare4}
\end{tcolorbox}

\begin{tcolorbox}[colframe=gray, colback=white, title=Evaluation Prompt-prompt for generated PR review, coltitle=white, colbacktitle=gray, fonttitle=\bfseries]

You are an expert software engineer performing a code review on a pull request. You are given:

\hspace*{2em} - The **pull request metadata**, including title and description

\hspace*{2em} - The **full original version of the modified file(s)** (before changes)

\hspace*{2em} - The **generated code** (generated code introduced in the PR.)

Your job is to provide **precise, objective, and actionable review comments** based on the changes introduced in the pull request. Use the **full original source** for understanding context and identifying problems.

Once you have identified issues in the code, return them as a JSON list, enclosed in backticks with JSON syntax highlighting. 

If you find the pull request is 100\% correct, output: "No comment."

pr meta data:
<pr\_metadata>

Original code:
<original\_file>

LLM-generated code:
<generated\_code>

\label{compare5}
\end{tcolorbox}

\begin{tcolorbox}[colframe=gray, colback=white, title=Evaluation Prompt-prompt for checklist score generation, coltitle=white, colbacktitle=gray, fonttitle=\bfseries]

You are given a checklist (a Python list of strings) and a code review (a string).

Your task is to determine how many checklist items are mentioned or addressed in the code review.

Return only a single number — the count of checklist items that are matched in the review.

if the checklist is: 'No checklist':

   \hspace*{2em} Your task is to determine whether the code review truely suggestion something.
   
   \hspace*{2em} Return only a single number.

  \hspace*{2em}  - If the code review contains any meaningful suggestion or feedback, return 0.
  
  \hspace*{2em}  - If the code review implies there's nothing to check, return 1.
  
  \hspace*{2em}  - Ignore the purpose, change and other infomation. focus only on the suggestions, if there is no suggestion/blank suggestion, return 1.

Checklist: {PR review checklist}

Code review: {Generated PR review}

**Only output the number.**

\label{compare6}
\end{tcolorbox}

\section{More examples}

We provide additional examples in the following content to illustrate how our prompts and evaluation framework operate in practice. These examples cover typical outputs from different stages of the pipeline, such as PR review comments, checklist generation, GPT model responses, and refined reviews produced by the Sphinx system.

\begin{tcolorbox}[colframe=gray, colback=white, title=PR review, coltitle=white, colbacktitle=gray, fonttitle=\bfseries]

In the SqlRegistryConfig class (around its field definitions), consider leveraging pydantic's StrictStr (as in the reference version) instead of plain str. This change enforces stricter type checking and helps catch configuration errors earlier. It would improve documentation and maintainability to add descriptive docstrings for the configuration fields (registry\_type and path) in SqlRegistryConfig as demonstrated in the reference implementation.

\label{exp1}
\end{tcolorbox}

\begin{tcolorbox}[colframe=gray, colback=white, title=PR review checklist, coltitle=white, colbacktitle=gray, fonttitle=\bfseries]

1. Update SqlRegistryConfig field definitions to use pydantic's StrictStr instead of plain str (enforces stricter type checking and catches configuration errors earlier)

2. Add descriptive docstrings for the registry\_type and path configuration fields in SqlRegistryConfig (improves documentation and maintainability)"

\label{exp2}
\end{tcolorbox}

\begin{tcolorbox}[colframe=gray, colback=white, title=GPT generated review, coltitle=white, colbacktitle=gray, fonttitle=\bfseries]

PR Overview

This pull request addresses a bug related to the use of the SQL Registry in the Feast project. The changes aim to enhance the configuration handling for SQL-based registries.

- Introduced a new `SqlRegistryConfig` class that extends `RegistryConfig` with additional fields specific to SQL registries.

- Updated the `SqlRegistry` class to accept the new `SqlRegistryConfig` type, allowing for more specific configuration options.

\label{exp4}
\end{tcolorbox}

\begin{tcolorbox}[colframe=gray, colback=white, title=Sphinx-GPT generated review, coltitle=white, colbacktitle=gray, fonttitle=\bfseries]

In the generated version, the SqlRegistryConfig class is defined as a simple dataclass with a single field (path: str). In the reference version, SqlRegistryConfig extends RegistryConfig and uses Pydantic StrictStr for type enforcement. Consider aligning with the reference by subclassing RegistryConfig and using StrictStr for the path field to ensure stricter type validation and consistency with the rest of the codebase.

The generated version SqlRegistryConfig does not include the registry\_type field. Adding this field (and its associated docstring) would improve clarity and maintain consistency with the reference implementation.

The generated version \_\_init\_\_ method for SqlRegistry accepts only a RegistryConfig parameter, whereas the reference version allows for an Optional Union RegistryConfig, SqlRegistryConfig. To improve flexibility and future compatibility, consider updating the \_\_init\_\_ signature to accept both types as in the reference version.

The generated version omits the docstring comments present in the reference version for the SqlRegistryConfig fields. Adding these docstrings would enhance code readability and maintainability.

Overall, while the core logic is functionally equivalent, these adjustments in configuration handling and type enforcement would bring the generated version closer to the reference implementation robustness and clarity.

\label{exp3}
\end{tcolorbox}


